# Tightly bound excitons in monolayer WSe$_2$

Keliang He[1], Nardeep Kumar[2], Liang Zhao[1], Zefang Wang[1], Kin Fai Mak[3], Hui Zhao[2], Jie Shan[1]

[1]Department of Physics, Case Western Reserve University, 10900 Euclid Avenue, Cleveland, Ohio 44106, USA
[2]Department of Physics and Astronomy, The University of Kansas, Lawrence, Kansas 66045, USA
[3]Kavli Institute at Cornell for Nanoscale Science and Laboratory of Atomic and Solid State Physics, Cornell University, Ithaca, New York 14853, USA

## Abstract

Exciton binding energy and excited states in monolayers of tungsten diselenide (WSe$_2$) are investigated using the combined linear absorption and two-photon photoluminescence excitation spectroscopy. The exciton binding energy is determined to be 0.37eV, which is about an order of magnitude larger than that in III-V semiconductor quantum wells and renders the exciton excited states observable even at room temperature. The exciton excitation spectrum with both experimentally determined one- and two-photon active states is distinct from the simple two-dimensional (2D) hydrogenic model. This result reveals significantly reduced and nonlocal dielectric screening of Coulomb interactions in 2D semiconductors. The observed large exciton binding energy will also have a significant impact on next-generation photonics and optoelectronics applications based on 2D atomic crystals.

One of the most distinctive features of electrons in two-dimensional (2D) semiconductors, such as single atomic layers of group-VI transition metal dichalcogenides (TMDs) [1], is the significantly reduced dielectric screening of Coulomb interactions. An important consequence of strong Coulomb interactions is the formation of tightly bound excitons. Indeed, recent theoretical studies have predicted a large exciton binding energy between 0.5 and 1eV in $MoS_2$ monolayers [2-10], a representative 2D direct gap semiconductor from the family of TMDs [11, 12]. These values for the exciton binding energy are more than an order of magnitude larger than that in conventional III-V based quasi-2D semiconductor quantum wells (QWs) [13, 14]. Such tightly bound excitons are expected to not only dominate the optical response, but also play a defining role in the optoelectronic processes, such as photoconduction and photocurrent generation in 2D semiconductors [1, 15]. On the other hand, little is known about these tightly bound excitons from the experimental standpoint, except the energy of the lowest energy one-photon active exciton states [11] and an indirect evidence of large binding energies through recent studies on trions, quasiparticles of two electrons and a hole or two holes and an electron [16-18]. Furthermore, a non-Rydberg series has been predicted for excitons in 2D semiconductors, arisen from the nonlocal character of screening of the Coulomb interactions [4, 19]. While a Rydberg series for the exciton energy spectrum has been observed in *bulk* $MoS_2$ [20, 21], similar experimental studies on monolayers of $MoS_2$ or other TMDs have not been reported [22].

The challenge in experimental determination of the exciton binding energy in 2D TMDs by linear optical methods, commonly used for bulk semiconductors [23] or conventional semiconductor QWs [13], lies in the identification of the onset of band-to-band transitions in the optical absorption or emission spectrum. Such an onset of band-to-band transitions has not been observed in 2D TMDs presumably due to the significant transfer of oscillator strengths from the band-to-band transitions to the fundamental exciton states, lifetime broadening, and potential overlap in energy with exciton states originated from higher energy bands and/or different parts of the Brillouin zone [11]. An alternative is to determine the exciton excited states and evaluate the binding energy from the level spacing based on a model. In the simple 2D hydrogenic model [24], where an electron-hole (*e-h*) pair in 2D interacts through a Coulomb potential, the energy spectrum is known as the Rydberg series $E_n = -\frac{E_b}{(n-1/2)^2}$ with an exciton binding energy $4E_b$. Each state *n* (=1, 2, 3…) is degenerate with angular momentum $l = 0, \pm 1, ..., \pm(n-1)$. For instance, the 2*s* (*l* = 0, one-photon allowed) and 2*p* ($l = \pm 1$, two-photon allowed) state are degenerate, lying at 8/9 of the exciton binding energy above the lowest energy 1*s* state. Measurements of the 1*s* and 2*s*/2*p* state allow the determination of the exciton binding energy in the 2D hydrogenic model.

In this Letter, we report a combined linear and nonlinear optical study on the exciton excited states and binding energy in monolayers of WSe$_2$, a 2D direct gap semiconductor from the family of TMDs, with optical and electronic properties similar to MoS$_2$. Our linear absorption measurement reveals up to five *s*-states from the A exciton series even at room temperature. Two-photon photoluminescence (2PPL) excitation spectroscopy [25-27] is employed to probe the *p*-states and measure the band edge energy directly. A band gap energy of 2.02eV and an exciton binding energy of 0.37eV have been determined for monolayer WSe$_2$ from the experimental results without relying on any specific exciton models. Further, the measured exciton excitation spectrum with much more evenly spaced energy levels is very distinct from the simple 2D hydrogenic model. This behavior can be qualitatively understood as a consequence of the nonlocal character of dielectric screening in 2D [4, 19].  Our experiment thus directly verifies the importance of Coulomb interactions and excitonic effects in 2D semiconductors.  The unique spectrum of exciton states with differing optical activities revealed by our experiment also presents new opportunities for the study and control of the spin/valley polarization in 2D TMDs through inter- and intra-excitonic processes [28-33].

In our experiment, atomically thin $WSe_2$ samples were mechanically exfoliated from their bulk form (2D semiconductors) onto Si substrates covered with a 100nm or 300nm $SiO_2$ layer or fused quartz substrates.  Monolayer samples were first identified according to their optical contrast with the substrate (Fig. 1a), and then confirmed by photoluminescence (PL) spectroscopy (Fig. 1b).  The PL was excited with a continuous wave (cw) HeNe laser at 1.96eV and recorded with a grating spectrometer equipped with either a liquid nitrogen or thermoelectrically cooled CCD camera.  A single narrow peak at ~1.65eV at room temperature (corresponding to the lowest energy exciton state A) with no lower-energy indirect gap emission features confirms the monolayer thickness [34, 35].

To probe the one-photon active exciton states, the linear absorption spectrum of monolayer $WSe_2$ was measured through the reflection contrast using broadband radiation from a super-continuum laser as described elsewhere [11, 36].  In short, the laser beam was focused onto the samples with a 50x microscope objective to a spot size of ~2μm.  Typically several hundred spectra of reflection contrast were averaged to improve the signal-to-noise ratio.  A typical spectral resolution is ~0.3meV.

To access the two-photon active exciton states in monolayer WSe$_2$, femtosecond infrared (IR) pulses in the energy range of 0.85-1.1eV generated from an optical parametric oscillator pumped by a Ti:sapphire laser were employed. The IR excited PL via two-photon absorption, instead of the direct attenuation of the IR excitation beam, was measured for higher detection sensitivity. The IR pulses were ~100fs in duration with a repetition rate of ~80MHz. They were focused onto the samples by a 20x IR objective to a spot size of ~2μm under normal incidence. The back-scattered signal was collected by the same objective and sent to a spectrometer after appropriate filtering. The energy of the IR source was tuned with a step size of ~10meV to obtain the 2PPL excitation spectrum. The spectral resolution is ~20meV determined by the bandwidth of the IR excitation pulses. To calibrate the variations in the IR pulse duration and beam size while changing its energy, we simultaneously measure the second-harmonic generation (SHG) from a z-cut single crystal quartz plate as a reference (see below). An excitation power below 2mW and 100$\mu$W, respectively, has been employed for the nonlinear and linear absorption measurements to avoid heating and radiation damage of the samples. No observable changes for both the PL spectral shape and quantum yield throughout the entire measurement on any sample were observed.

Figure 2a illustrates the linear absorption spectrum of monolayer WSe$_2$ on fused quartz (red line). In the energy range of 1.5-2.3eV, there are two prominent exciton peaks at 1.65 and 2.08eV, respectively. These peaks, labeled A and B, correspond to the lowest energy exciton states originated from transitions from the two highest energy spin-orbit split-off valence bands to the lowest energy conduction bands around the K(K') point in the Brillouin zone [34]. The large energy separation between the A and B exciton state (~ 0.43eV) due to strong spin-orbit coupling in $WSe_2$ opens up a window for the observation of exciton excited states of the A series. Below we will focus only on the A exciton series.

A careful examination of the linear absorption spectrum reveals resonance A' at 1.82eV, about 0.16eV above the prominent A peak. Furthermore, three additional resonances at higher energies with increasingly smaller oscillator strengths (marked with *) can be identified as dips from the second-order numerical derivative of the absorption spectrum [23] (Fig. 2b). These features have been observed in all five samples studied in this experiment. The dip immediately above the A' energy (with an amplitude smaller than the next identified resonance) was not reproduced and likely an artifact. For temperature <150K, a $6^{th}$ resonance can also be identified. Further, all these identified resonance features show a similar blue shift with temperature as the A peak, dictated primarily by temperature renormalization of the band gap [15, 23]. (See Supplemental Material S2 for details on the experimental results and their analysis.) Based on these evidences we assign these features as one-photon active states from the A exciton series and label them in order of increasing energy as 1$s$, 2$s$ ... in analogy to the hydrogenic Rydberg series. We note that first, the large level spacing of $E_{5s} - E_{1s}$ ~0.3eV suggests that the exciton binding energy is at least 0.3eV. Second, we did not observe any noticeable dependence of the states on substrate studied in this experiment. And third, while the 1$s$ state narrows from ~40meV at room temperature to ~10meV at 30K (Fig. S4), consistent with an earlier PL measurement on the same material [33], the width of the 2$s$ and 3$s$ state (~30meV) does not depend on temperature. The latter is a clear evidence of significant lifetime broadening from effects such as energy-dependent exciton-phonon scattering. Given the significant lifetime broadening for the exciton excited states, we will focus on room temperature in the nonlinear optical study below.

Now we turn to the discussion of the two-photon allowed exciton states. In figure 1c, we show the emission spectrum of monolayer WSe$_2$ in the visible range under excitation of an IR pulse of energy $\hbar\omega$ = 1.07eV. The spectrum consists of a narrow peak at $2\hbar\omega$ (blue line), corresponding to the SHG from monolayer WSe$_2$ [35], and a weaker feature peaked at 1.65eV (red line). The latter matches the PL spectrum of the sample under cw excitation (green line) and depends on the excitation power quadratically as the second-harmonic (SH) signal (Fig. 1d). It is thus confirmed that PL can indeed be induced through two-photon absorption and will be used to characterize the two-photon absorbance.

Two-photon absorption in a sample can be described as a third-order nonlinear process, and the two-photon absorbance, by the imaginary part of the third-order nonlinear susceptibility $Im[\chi_S^{(3)}(\omega,-\omega,\omega)]$. To extract this parameter, we normalize the 2PPL intensity $I_{2PPL,S}$ from the sample by the SH intensity detected from a reference, $I_{2\omega,Q}$, under identical experimental conditions. Z-cut single crystal quartz was chosen as the reference since both the fundamental and SH frequency for the energy range studied here are far away from the quartz band gap, and the dispersion in both the linear and nonlinear optical properties can be ignored [37]. If we assume the PL quantum yield independent of the IR excitation energy, the third-order nonlinear sheet susceptibility of a monolayer sample on a substrate can be derived as [38]

$$Im[\chi_S^{(3)}(\omega,-\omega,\omega)] \propto \omega^{-1}|L_\omega^{-4}|\frac{I_{2PPL,S}}{I_{2\omega,Q}}. \qquad (1)$$

Here $L_\omega$ is the local field factor at the fundamental frequency, which converts the incident excitation field to the field in the sample, and the factor $\omega^{-1}$ arises from processes at the surface/interface. For the experimental geometry of monolayer samples on Si covered with a SiO$_2$ layer of thickness $L_{SiO2}$, the local field factor is given as $L_\omega = \frac{t_{12}}{1-r_{23}r_{21}e^{-2i\varphi}}$ with Fresnel coefficients $t_{12} = \frac{2}{1+n_{SiO2}}$, $r_{21} = \frac{n_{SiO2}-1}{n_{SiO2}+1}$, $r_{23} = \frac{n_{SiO2}-n_{Si}}{n_{SiO2}+n_{Si}}$, and phase shift $\varphi = n_{SiO2}\omega L_{SiO2}/c$ from propagation in the SiO$_2$ layer.

Figure 3a illustrates the normalized 2PPL spectra $\frac{I_{2PPL,S}}{I_{2\omega,Q}}$ of a monolayer sample of WSe$_2$ on Si with a 300nm SiO$_2$ layer at varying IR excitation energies $\hbar\omega$ = 0.85-1.1eV at room temperature. No changes are observable in the PL spectral shape. The third-order sheet susceptibility $Im[\chi_S^{(3)}]$ is then obtained from the integrated PL intensity (1.66 - 1.69eV) for each excitation energy according to Eq. (1). The result is plotted against $2\hbar\omega$ in figure 3b and for comparison, also in the same plot for the linear absorption spectrum (blue symbols, Fig. 2a).

$Im[\chi_S^{(3)}]$ shows a non-monotonic dependence on the excitation energy with three interesting features. (i) No signal can be measured for $2\hbar\omega < 1.72$eV. This indicates that the A exciton state is strongly suppressed, which is consistent with its assignment as a 1s state. (ii) $Im[\chi_S^{(3)}]$ increases rapidly for $2\hbar\omega > 1.8$eV and forms a broad peak A". Although the limited spectral resolution and signal-to-noise ratio of this measurement does not allow us to resolve any sub-features of the broad peak A", a careful comparison of the two-photon and one-photon absorption spectrum (Fig. 2a) reveals consistent absorption enhancement around the energies of the 2s, 3s and 4s states, which suggests that the broad A" peak is likely a superposition of the corresponding np states. This assignment is consistent with the selection rules for the excitation pulse polarized in the plane of the sample [39]. On the other hand, however, it is unclear why the 2*p* state has smaller two-photon absorbance than 3*p* or 4*p*. Future theoretical and experimental studies on the nature and assignment of the exciton states are warranted. (iii) $Im[\chi_S^{(3)}]$ drops to a value of about half of its peak followed by a weak upward trend for $2\hbar\omega > 2$eV. This feature is compatible with band-to-band transitions. In the simple two parabolic band model including the excitonic effect, the two-photon absorption transition rate of the band-to-band transitions scales linearly with the two-photon energy $\sim 1 + \frac{2\hbar\omega - E_g}{4E_b}$ when $2\hbar\omega$ is above the band gap energy $E_g$ [39]. We describe the experimental 2PPL excitation spectrum (symbols, Fig. 3b) by the sum (solid green line) of a Gaussian function (dotted red line), which qualitatively accounts for the total contribution of all *p*-states, and a linear function with a step at the band gap energy (dotted red line). A good agreement is obtained for $E_g$ = 2.02eV with a broadening of 80meV. We thus determine the A exciton binding energy in monolayer WSe$_2$ to be $E_g - E_{1s}$=0.37eV.

Finally, we would like to understand the origin of the non-Rydberg exciton series observed in monolayer WSe$_2$. In the energy diagram of figure 4, we represent the *s*-states by red lines at their peak energies (one-photon active), and the broad A" state by a blue box (two-photon active). In the right panel we compare it to the 2D hydrogenic model, in which the bottom of the continuum and the exciton binding energy have been assumed to be the same as in the experiment (2.02 and 0.37eV, respectively). The experimental states are clearly much more evenly spaced than predicted by the 2D hydrogenic model. For instance, the 1*s* and 2*s* splitting contributes to <1/2 instead of the predicted 8/9 of the total exciton binding energy [19, 24]. Such a behavior can be qualitatively understood by considering the problem of dielectric screening in 2D [19]. Dielectric screening in a 2D semiconductor is significantly reduced compared to its 3D analog since the material is polarizable only in the plane. This effect explains the large level spacing and binding energies of the excitons in monolayer $WSe_2$. Further, as a result of the induced in-plane polarization, two point charges living in a 2D plane interact effectively as two thin rods with charges decaying into the out-of-plane direction [19]. Thus at large distances the interaction behaves as an unscreened Coulomb potential, but at small distances diverges logarithmically, resulting in a weaker interaction potential. Such nonlocal screening affects the low-energy states the most (by lifting their energies) because of their small radii and results in a non-Rydberg series. We note that while a non-Rydberg series could also arise from non-parabolic band dispersion, this factor does not play any significant role in monolayer $WSe_2$. Both the conduction and valence bands near the K(K') point can be well described by parabolic dispersion [9, 40]. With increasing energies, the states are expected to be more Rydberg-like and provide better basis for the estimation of the exciton binding energy based on the 2D hydrogenic model. For instance, we obtain a binding energy of 0.33eV from the 4$s$ and 5$s$ state, which is fully compatible with the value determined from 2PPL. (See Supplemental Material S3 for details).

In conclusion, we have directly probed, by complementary linear and nonlinear optical methods, the exciton excitation spectrum and the band gap in $WSe_2$ monolayers. The tightly bound excitons in this material allow us to observe up to five exciton states in linear absorption even at room temperature. The 2PPL excitation spectroscopy determines the band gap energy to be 2.02eV, revealing a large exciton binding energy of 0.37eV. These results are distinct from the predictions of the simple 2D hydrogenic model that ignores screening. Since optically active excitons play a central role in most optoelectronic processes, the observed large exciton binding energy and exciton excitation spectrum will form a basis for future understanding and optimization of optoelectronic devices based on 2D semiconductors.

## Acknowledgements

This work was supported by the Research Corporation Scialog Program and the National Science Foundation grant DMR-0349201 and 0907477 at Case Western Reserve University and the National Science Foundation grant DMR-0954486 at the University of Kansas.

**Figure captions**

**Figure 1.** (a) Optical reflection image of WSe$_2$ flakes on a Si substrate covered by a 300nm SiO$_2$ layer. A monolayer sample (middle) is outlined by dashed blue lines. (b) Photoluminescence (PL) spectrum of monolayer WSe$_2$ excited by a cw HeNe laser at 1.96eV. (c) Emission spectrum of monolayer WSe$_2$ under the excitation of femtosecond infrared pulses centered at 1.07eV. It consists of two features corresponding to the second-harmonic generation (SHG) (blue) and two-photon PL (red). The latter is magnified by 15 times. The PL excited by the cw HeNe laser (green) is included for comparison. (d) Excitation power dependence of the integrated SHG and two-photon PL.

**Figure 2.** (a) Linear absorption (red line, right axis) and 2PPL excitation spectrum (blue symbols, left axis) measured on monolayer WSe$_2$ at room temperature. Each data point of the 2PPL excitation spectrum corresponds to an integrated PL (1.664 - 1.687eV) normalized by the reference SHG signal from a z-cut quartz crystal according to Eq. (1). The uncertainty corresponds to the spectral resolution, determined by the bandwidth of the excitation pulse. A and B correspond to the fundamental exciton resonances arisen from transitions from the two highest energy spin-orbit split-off valance bands and the lowest energy conduction bands at the K(K') point of the Brillouin zone. A' and A'' denote the 2$s$ state and a broad $p$-peak observed in one-photon and two-photon absorption, respectively. (b) Second-order numerical derivative of the linear absorption spectrum of (a). Black dashed line denotes the band edge energy of 2.02eV determined from the fit described in the text.

**Figure 3.** (a) Representative PL spectra of monolayer WSe$_2$ excited by femtosecond IR pulses centered at 0.85-1.1eV. The spectra were normalized by the SHG intensity from a z-cut single crystal quartz plate recorded under identical experimental conditions. (b) Experimental 2PPL excitation spectrum (symbols) and fit (green line) including contributions from both excitons and band-to-band transitions (dotted red lines) as described in the text.

**Figure 4.** Exciton excitation spectrum of monolayer WSe$_2$ determined experimentally in this work (left panel) is compared with the 2D hydrogenic model (right panel). Red lines denote the one-photon active states and the blue box is the unresolved two-photon active states. The exciton binding energy and the bottom of the continuum in the 2D hydrogenic model are chosen to match the values obtained from experiment.

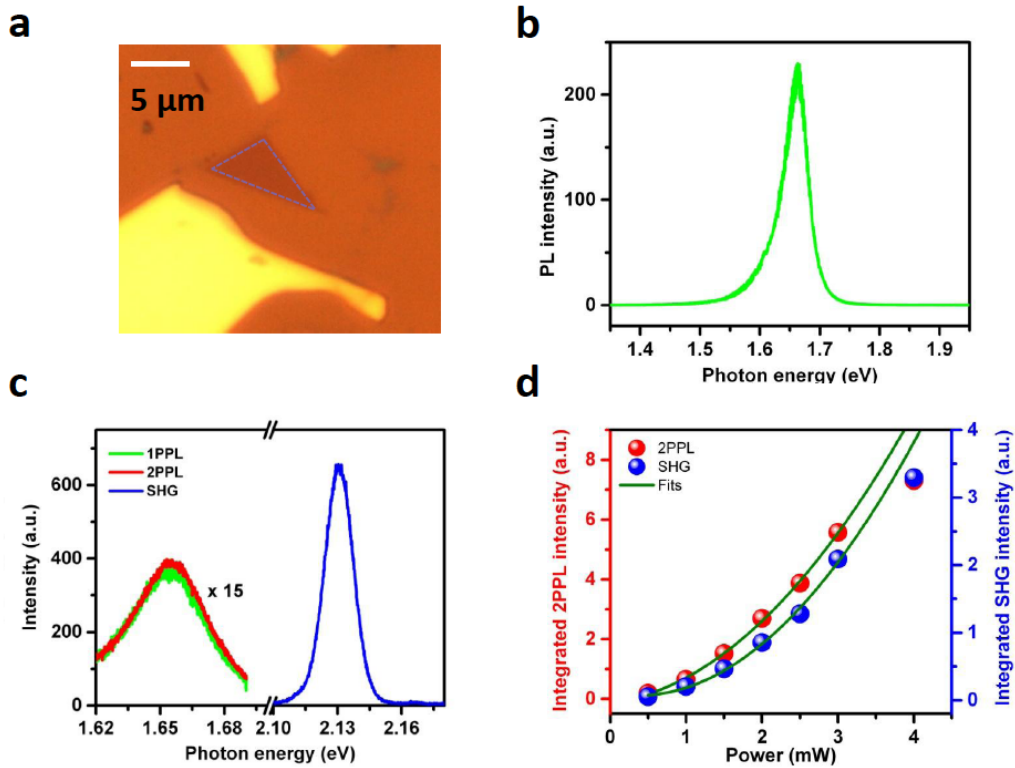

**Figure 1.** (a) Optical reflection image of $WSe_2$ flakes on a Si substrate covered by a 300nm $SiO_2$ layer. A monolayer sample (middle) is outlined by dashed blue lines. (b) Photoluminescence (PL) spectrum of monolayer $WSe_2$ excited by a cw HeNe laser at 1.96eV. (c) Emission spectrum of monolayer $WSe_2$ under the excitation of femtosecond infrared pulses centered at 1.07eV. It consists of two features corresponding to the second-harmonic generation (SHG) (blue) and two-photon PL (red). The latter is magnified by 15 times. The PL excited by the cw HeNe laser (green) is included for comparison. (d) Excitation power dependence of the integrated SHG and two-photon PL.

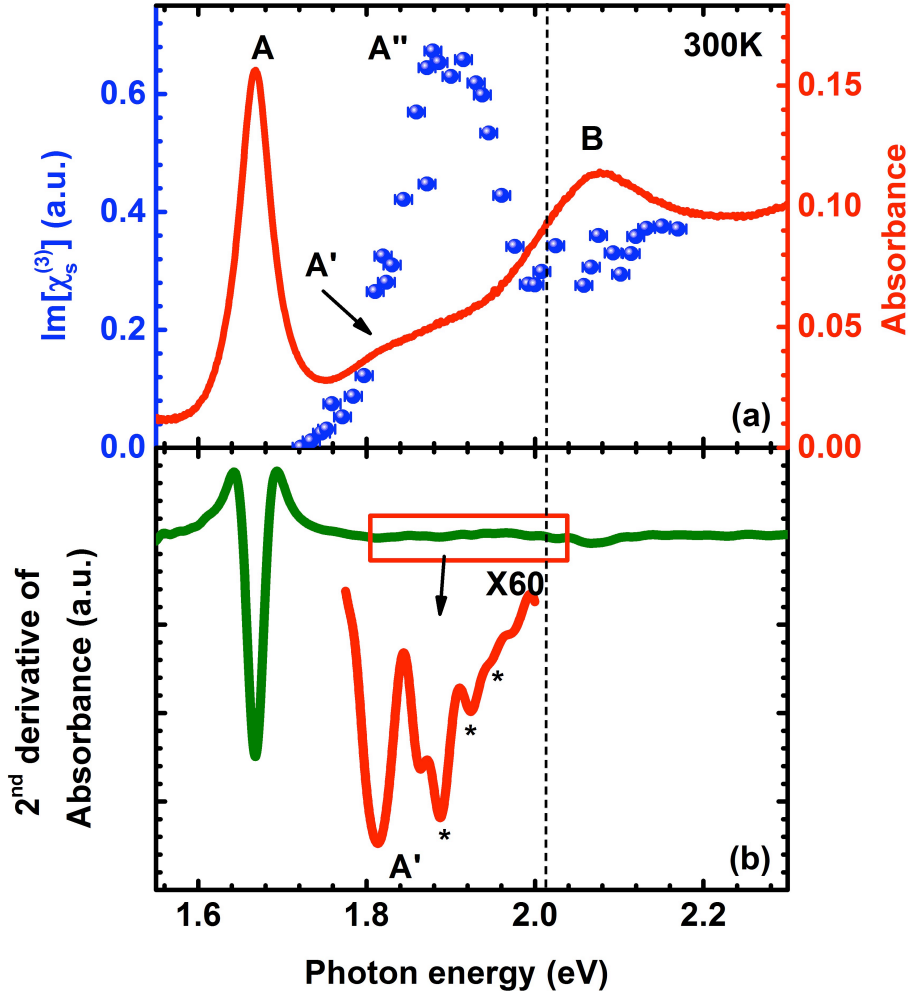

**Figure 2.** (a) Linear absorption (red line, right axis) and 2PPL excitation spectrum (blue symbols, left axis) measured on monolayer $WSe_2$ at room temperature. Each data point of the 2PPL excitation spectrum corresponds to an integrated PL (1.664 - 1.687eV) normalized by the reference SHG signal from a z-cut quartz crystal according to Eq. (1). The uncertainty corresponds to the spectral resolution, determined by the bandwidth of the excitation pulse. A and B correspond to the fundamental exciton resonances arisen from transitions from the two highest energy spin-orbit split-off valance bands and the lowest energy conduction bands at the K(K') point of the Brillouin zone. A' and A" denote the 2s state and a broad p-peak observed in one-photon and two-photon absorption, respectively. (b) Second-order numerical derivative of the linear absorption spectrum of (a). Black dashed line denotes the band edge energy of 2.02eV determined from the fit described in the text.

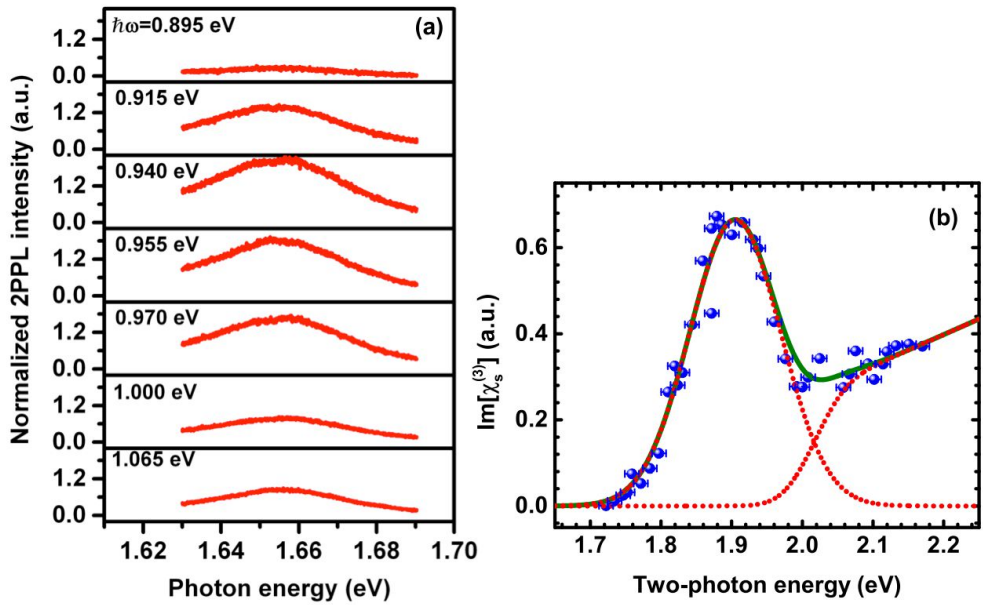

**Figure 3.** (a) Representative PL spectra of monolayer WSe$_2$ excited by femtosecond IR pulses centered at 0.85-1.1eV. The spectra were normalized by the SHG intensity from a z-cut single crystal quartz plate recorded under identical experimental conditions. (b) Experimental 2PPL excitation spectrum (symbols) and fit (green line) including contributions from both excitons and band-to-band transitions (dotted red lines) as described in the text.

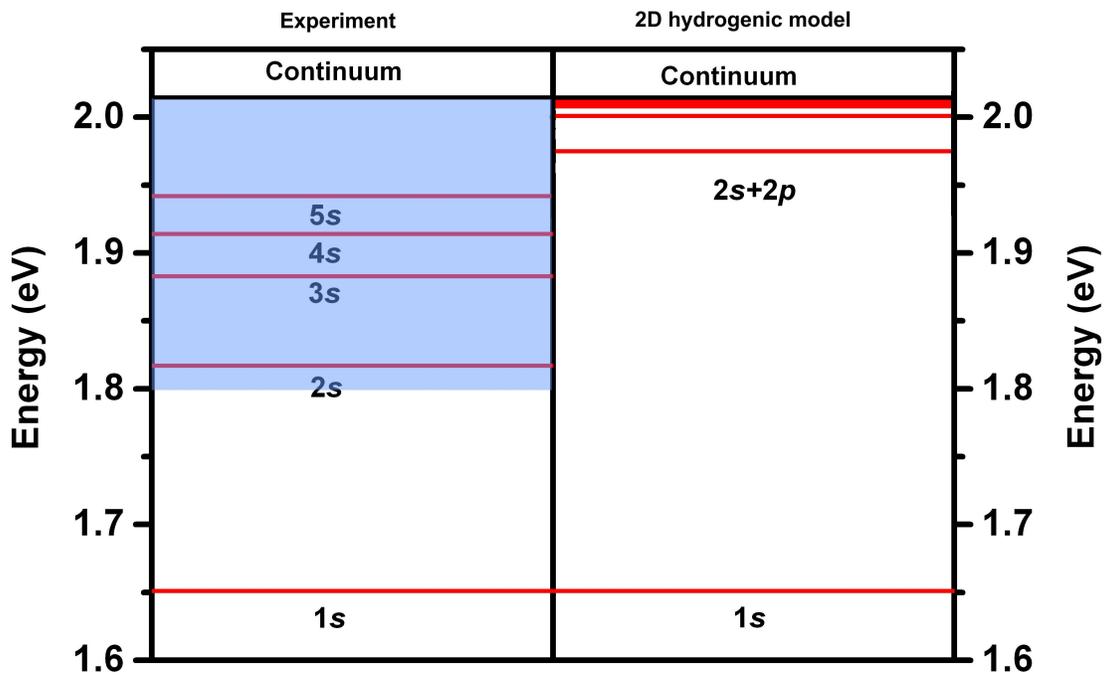

**Figure 4.** Exciton excitation spectrum of monolayer WSe$_2$ determined experimentally in this work (left panel) is compared with the 2D hydrogenic model (right panel). Red lines denote the one-photon active states and the blue box is the unresolved two-photon active states. The exciton binding energy and the bottom of the continuum in the 2D hydrogenic model are chosen to match the values obtained from experiment.